**b) Level crossing in AFM rings**

In a AFM ring of the kind discussed above with a S=0 singlet ground state and a S=1 triplet excited state, an external magnetic field **H** removes the residual Kramers degeneracy of the triplet state and induces multiple level-crossings (LCs) at specific magnetic field values $H_{ci}$, whereupon the ground state of the molecule changes from $S = 0$ to $S = 1$ ($H_{c1}$), from $S = 1$ to $S = 2$ ($H_{c2}$), and so on. Because of magnetic anisotropy, the values of $H_{ci}$ depend on the angle between **H** and the molecular axis **z** [15,35]. The situation of near-degeneracy of the magnetic levels near the LC fields, and particularly the situation of level repulsion or anticrossing (LAC), is favorable to the observation of quantum tunneling and quantum coherence [36]. A situation of LC is also found between singlet and triplet states in 1D gapped quantum magnets [37], but the physical context and the continuum of excited states make the situation not comparable to that in finite-size magnets. For the spin dynamics near LC, a crucial issue is the role played by the coupling between magnetic molecular levels and the environment such as phonons and/or nuclear spins [38]. Essential information on this problem are accessed through measurements of the nuclear spin-lattice relaxation rate $1/T_1$ since the nuclei probe the fluctuations of the local field induced at the nuclear site by the magnetic moments of the transition ions. The spin dynamics at LC was studied by means of $^1$H NMR in Fe10 powder sample [39], Fe6:Li(BPh4) single crystal [40] and Cr8 single crystal [41] close to $H_{c1}$ and $H_{c2}$. We will review in the following the main results and conclusions derived in the above mentioned studies.

One crucial issue in the study of LC in AFM rings is the structure of the magnetic levels at the critical field where the S=0 state becomes degenerate with the S=1, M=1 state (first level crossing) and similarly for the other LC's. If the magnetic Hamiltonian does not contain terms that admix the two degenerate levels one can have in principle a true LC. If on the other hand the Hamiltonian contains terms which strongly admix the levels one expects a gap at the critical field resulting in what we call level anticrossing (LAC). The spin Hamiltonian describing a $N$-membered ringlike spin topology can be written as:

$$H = J \Sigma_i \mathbf{s}_i \cdot \mathbf{s}_{i+1} + \Sigma_i U(\mathbf{s}_i) + \Sigma_{i>j} U_{i,j}(\mathbf{s}_i, \mathbf{s}_j) + g\mu_B \mathbf{H} \Sigma_i \mathbf{s}_i \qquad (22)$$

where $i$ and $j$ run from 1 to $N$. The first term is the nearest-neighbor Heisenberg exchange interaction, the second and third terms represent the crystal-field anisotropies and the anisotropic spin-spin interactions (including eventually a Dzyaloshinski-Moriya (DM) term $C_{ij} \mathbf{s}i \otimes \mathbf{s}j$), respectively, and the last one is the Zeeman term. In Eq. 22, $\mathbf{s}_{N+1} = \mathbf{s}_1$ and the exchange interactions are supposed to be limited to nearest neighbors. Magnetic torque and specific heat measurements have been performed as a function of magnetic field in molecular rings [15,35,40]. The former is a powerful tool to locate level crossings, which lead to abrupt variations of the torque signal at low temperature, and to determine the ZFS parameters ($D_S$) of excited spin states. Specific heat measurements, when performed at sufficiently low temperature, are in principle able to distinguish a true LC from a LAC [40]. Particularly interesting is the possibility to extract the value of the energy gap at the anticrossing field, which is directly proportional to the matrix element connecting the two states involved.

Measurements of proton spin-lattice relaxation as a function of the magnetic field at fixed T are shown in Figs. 15, 16 and 17 for the three AFM rings Fe10, Fe6(Li) and Cr8. In all three cases a strong enhancement of $T_1^{-1}$ is observed in correspondence to critical field values for LC (or LAC). The enhancement is clearly related either to cross relaxation effects or to magnetization fluctuations due to the almost degeneracy of the crossing magnetic levels as will be discussed further on. In Fe10 the peaks of $T_1^{-1}$ are at 4.7, 9.6 and 14 Tesla. The field dependence of the gap in Fe10 leading to the first LC for H parallel to the z axis should be $\Delta(H) = \Delta(0) + D_1/3 - g\mu_B H$ where the anisotropy parameter $D_1 = 3.2$ K and $\Delta(0) = 5.5$K from Table II. The gap closes for H= 6.3K/1.33 K/T ≈ 4.7T in excellent agreement with the NMR data. Since the measurements in Fe10 were performed in a powder sample there is a distribution of LC fields which generate a broadening of the $T_1^{-1}$ vs H curve. In Fe6(Li) single crystal the maximum of $T_1^{-1}$ can be observed at 11.7T in excellent agreement with torque measurements for the same angle θ~25° between the z axis and H. In Cr8 two peaks in $T_1^{-1}$ could be observed at 6.85T and 13.95T. The measurements were performed in single crystal but the orientation of the crystal is not defined. From the knowledge of $D_1$ = 2.3K and $\Delta(0)$=8.5 in Table II one deduce that the NMR experiments were performed at θ~65° (1st crossing) and θ~50° (2nd crossing).

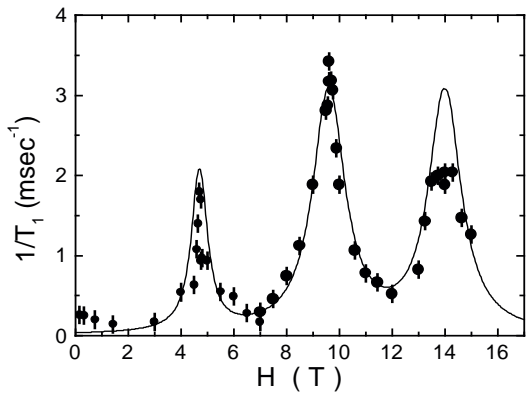

Fig. 15: Magnetic field dependence of $^1$H-1/$T_1$ in Fe10 powders measured at low temperature T≈1.5K.

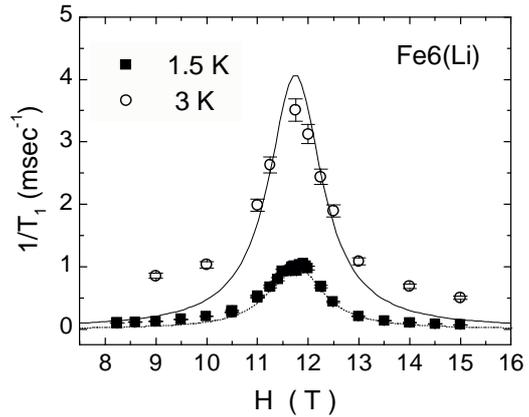

Fig. 16 : Magnetic field dependence of $^1$H-1/$T_1$ in Fe6(Li) single crystal measured at T=1.5 and 3K, for θ≈20° (θ = angle between H and the molecular axis).

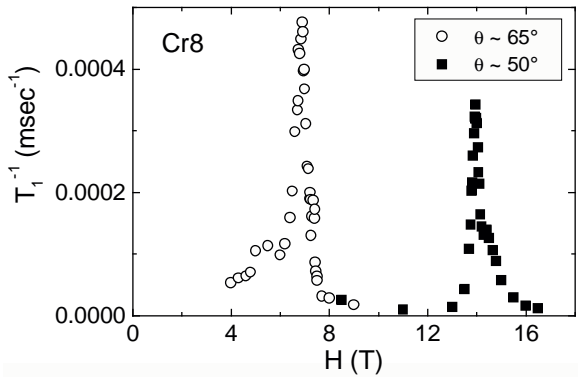

Fig. 17: Magnetic field dependence of $^1$H-1/$T_1$ in Cr8 single crystal measured at T=1.5K, for two slightly different values of the angle θ between H and the molecular axis.

Having established that the peaks in the proton $T_1^{-1}$ vs H curve are indeed evidence for LC (or LAC) it remains to be seen what kind of information one can obtain about the spin dynamics at level crossing from NMR measurements. The processes involved in nuclear spin-lattice relaxation close to LC or LAC are currently not well understood. Ultimately, the challenge is to be able to distinguish between thermal and quantum spin fluctuations in the molecule. In the presence of true LC or very small LAC, there are two magnetic field values for which the level separation $\Delta_1 = g\mu_B|H_{ci}-H|$ matches the $^1H$ Zeeman energy $h\nu_L$. At these two field values, one expects a large enhancement of $1/T_1$ due to a direct exchange of energy between electronic and nuclear reservoirs, i.e. a cross relaxation [17,39]. The cross-relaxation depends on the relative values of the nuclear and electron spin-lattice relaxation rates $T_1$ and $T_1^{el}$ respectively, which limit the energy transfer to the "lattice", and on the rate of internal energy exchange between the nuclear Zeeman reservoir and the molecular magnet reservoir. If $T_1^{el}$ is much shorter than the nuclear $T_1$ and the exchange time between reservoirs is very short, one expects a large T-independent double-peak of $T_1^{-1}$ close to LC. On the other hand, if $T_1^{el} \geq T_1$, the cross-relaxation may not be effective. However, one should keep in mind that in practice this situation can only occur in the case of ultra small LACs (as small as the nuclear Zeeman energy, i.e. of the order of mK) and that the separation between the two peaks of $1/T_1(H)$ would also be as small as the nuclear Zeeman energy. Furthermore, to be experimentally observable as two separate peaks with finite width, this mechanism would require some broadening of the levels. If the LAC is large (with respect to nuclear energies), the nuclear relaxation should have dominant contributions from indirect processes (e.g. Raman-like, Orbach-like [17, 38]). The width of the peak and its T dependence may be influenced by many parameters, and should certainly be different from the above-cited case of cross-relaxation. To obtain a phenomenological expression for $1/T_1$ near the LC fields, we suppose that the fluctuations of the magnetization between two adjacent magnetic states will drive the nuclear relaxation, that can be written as:

$$1/T_1 = A^2 J(\omega_L) \qquad (23)$$

where $J(\omega_L)$ is the spectral density of spin fluctuations at the nuclear Larmor frequency $\omega_L$ and $A$ is an average hyperfine coupling constant. In a phenomenological model where the spin-spin correlation function is supposed to decay exponentially with time, we assume

$$1/T_1(T,H) = A^2 \Gamma/[\Gamma^2 + (\hbar\omega_L - \Delta(H))^2] \qquad (24)$$

where $\Gamma$ is a temperature-dependent damping factor associated with level broadening. In the case of nuclear relaxation induced by a purely quantum process, one does not expect any T dependence of $T_1$. The relaxation can still be described by an expression similar to Eq. 24, with a T-independent $\Gamma$, now having a different meaning. Near the LC fields we can write

$$\Delta(H) = \{[g\mu_B(H_{ci}-H)]^2 + \Delta_i^2\}^{1/2} \qquad (25)$$

where $\Delta_i$ is the temperature independent "gap" at the anticrossing. The use of Eq. 24 near the LC condition is justified by the fact that the next excited state is at least several Kelvin higher in energy. Three different cases regarding the peak of $T_1^{-1}$ vs H can be distinguished :

a) for $\Delta_i \gg \Gamma(T)$, the width of the peak is determined by $\Delta_i$ while the intensity depends mainly on $\Gamma(T)$. Qualitatively, for $H = H_{ci}$ one has $1/T_1(T) \propto \Gamma(T)$. This situation corresponds to a well-identified LAC.

b) for $\Delta_i \ll \Gamma(T)$, both the width and the intensity of the peak are determined by $\Gamma(T)$, and for $H = H_{ci}$ one has $1/T_1(T) \propto 1/\Gamma(T)$. In this situation, it is not possible to identify a LAC, either because there is a true LC or because $\Delta_i$ is too small to influence NMR relaxation.

c) for $\Delta_i \sim \Gamma(T)$, the width of the peak is due to both $\Delta_i$ and $\Gamma(T)$.

Note that in general $T_1^{-1}(H = H_{ci})$ *vs.* T shows a maximum if $\Gamma(T) = |\Delta_i - \hbar\omega_L|$.

Let us now focus on the experimental results. As previously noted, in Fe10 three broad peaks were observed in the powder sample at the first three LC fields (see Fig. 15). In the range $1.5 < T < 4.2$K, it was shown that $1/T_1(T)$ = const, thus suggesting that $\Gamma$ is temperature independent. Out of the crossing field values, $1/T_1$ has an exponential behavior [39] (as also shown by the authors of ref. [32]). Since Fe10 data can be fitted by Eq. 24) assuming $\Delta_1 = 0$, the physical situation could correspond to the case discussed above at point b) , i.e. a true LC or small LAC. First principles $T_1^{-1}$ calculations confirmed this scenario [42].

In the case of Fe6(Li) the peak of $1/T_1$ vs H (see Fig. 16) is quite broad in spite of the fact that the measurements are done in single crystal. The NMR data appear to confirm the specific heat results which are suggestive of a LAC at $H_{c1}=11.8$ Tesla with a anticrossing gap $\Delta_1/k_B = 0.86$ K [40]. In fact the proton relaxation data can be fitted well near the maximum by using Eq. 24 with the above values for the parameters and with $A = 8.5 \cdot 10^7$ rad s$^{-1}$. The two curves in Fig. 16 at 1.5K and 3K are well reproduced by assuming a damping factor varying quadratically with temperature $\Gamma \sim T^2$ [40]. We note that the *T*-dependent broadening $\Gamma$ of the magnetic levels is small ($\Gamma = 0.26$ Tesla at 3 K) and thus it affects only the magnitude of $1/T_1$, while the width of the $1/T_1$ peak is determined by the anticrossing gap $\Delta_1$. This corresponds to the situation of the above-cited case (a). The observation of a large LAC has raised the problem of explaining its occurrence in terms of antisymmetric interaction in Eq. 22 [40, 43, 44].

In Cr8, two LCs are observed in the NMR (see Fig. 17). Both peaks show a high field shoulder whose origin is not yet understood and will be disregarded in the discussion which follows. The peaks of $1/T_1$ are narrower than in Fe6:Li(BPh4) by a factor ~2.5 for the first crossing and by a factor ~4 for the second crossing. Actually, the dipolar or hyperfine interaction extracted from $^1$H NMR spectra were estimated to yield a broadening of about 0.1÷0.2 T, not far from the measured width of the peak of $1/T_1$ on the second crossing. This suggests that LAC effects are smaller in Cr8 than in Fe6, or that the damping factor in Cr8 is comparable to or greater than the energy gap at the anticrossing. This experimental observation seems to be in qualitative agreement with specific heat data [45] that indicate anticrossing gap values much smaller than in Fe6(Li). However, from *C*(H) data it was not possible to obtain more than an upper limit for $\Delta_1$ (~0.2 K). As a consequence one has to determine all three parameters *A*, $\Delta_1$ and $\Gamma$ in Eq.

24 from $T_1$ data alone. This requires extensive $T_1$ measurements as a function of H and T not currently available. We note that the above discussion in terms of the phenomenological expression 24, is strictly valid only in the very vicinity of the LC critical fields. The field dependence at a fixed very low temperature value over a wide range of H on both sides of $H_{ci}$'s follows a behavior of the type $1/T_1 = A(H,T) \exp(-\Delta(H)/k_B T)$. The exponential dependence of $1/T_1$ with a field-dependent gap parameter $\Delta(H)$ reflects the same nuclear spin-lattice direct relaxation process described by Eq. 21 in the previous section

**c) Local spin configuration in the ground state of nanomagnets Mn12, Fe8**

As mentioned before at very low temperature ($\leq 4K$) both Mn12 and Fe8 are in their magnetic ground state and the magnetization of the molecule is frozen in the time scale of an NMR experiment leading to the possibility of observing zero field NMR. The $^{55}$Mn NMR spectrum in zero external field in Mn12 was first observed by Goto et al. [46]. A detailed study was later reported by us [47]. The $^{55}$Mn spectrum is shown in Fig. 18. The narrow low frequency line originates from the $Mn^{4+}$ ions while the two broader lines are from the $Mn^{3+}$ ions whereby the broadening of the latter two lines is of quadrupole origin [47, 48]. The $^{57}$Fe NMR signal in zero external field was detected in isotopically enriched Fe8 [49, 50] and the complete spectrum is shown in Fig. 19. The eight narrow lines correspond to the eight non equivalent Fe sites in the molecule. The identification of the eight lines with the corresponding Fe sites was possible from measurements as a function of the applied magnetic field [49, 50]. For protons the internal field is small and the NMR in zero field is weak and can be observed only at low frequency over a broad frequency interval (2-4MHz) due to the presence of many inequivalent proton sites in the molecule. We have observed the signal in zero field but the complete spectrum has never been published. On the other hand the proton NMR in an external magnetic field can be observed easily. It shows a broad structured spectrum with a field independent shift of the lines of order of the internal field. The proton spectrum in Fe8 at 1.5K is shown in Fig. 20 [25,51]. The proton spectrum in Mn12 at low temperature is shown in Fig. 21. Further details about the zero field NMR and about the analysis of the hyperfine field at the different nuclear sites are given in the refs. 47 and 48.

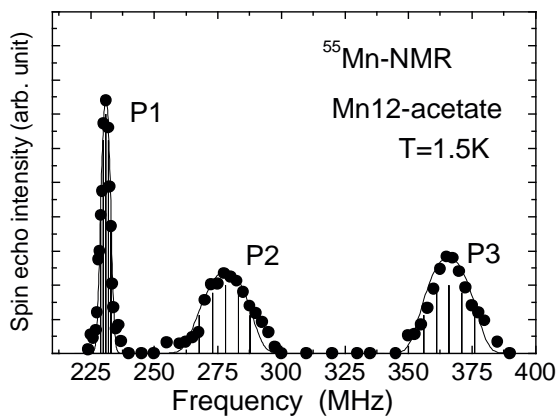

Fig. 18. $^{55}$Mn-NMR spectra in Mn12 oriented powders measured at T=1.5K under zero magnetic field.

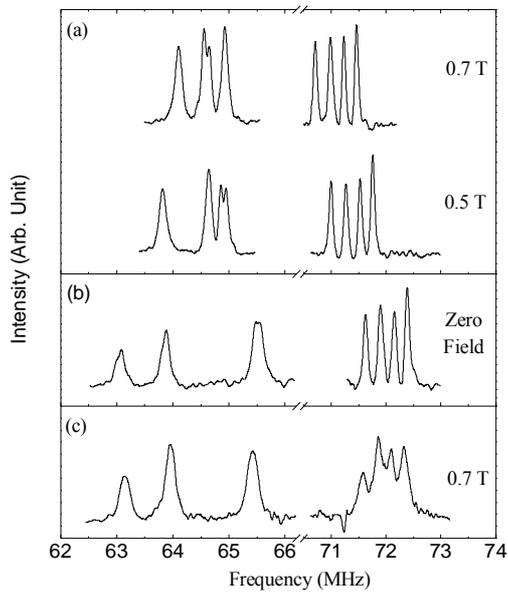

Fig. 19. $^{57}$Fe-NMR spectra on Fe8 powders measured at T=1.5K. (a) H=0.5 and 0.7T parallel to the easy axis (b) H=0. (c) H=0.7T perpendicular to the easy-axis.

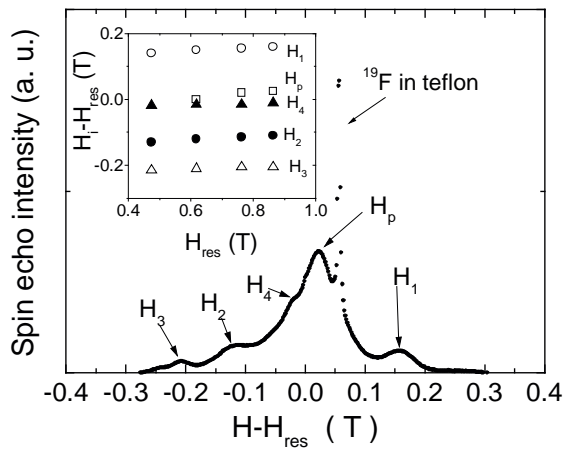

Fig. 20: $^{1}$H-NMR spectrum of Fe8 oriented powders at 36.8MHz ($H_{res}$=0.864T) and T=1.4K. In the inset the shifts of the lines at different fields $H_{1...4,p}$ as a function of $H_{res}$ are shown to be constant. Here H is parallel to the easy-axis.

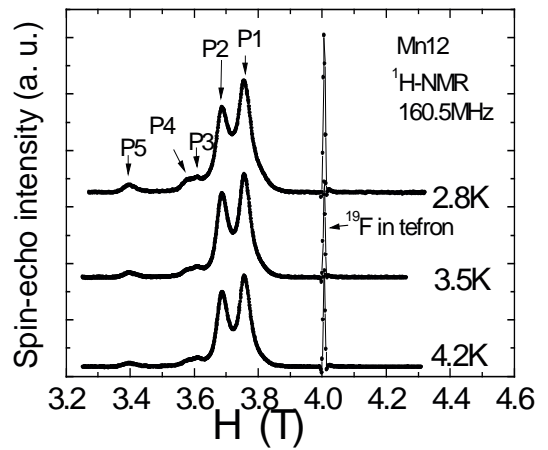

Fig. 21: $^{1}$H-NMR spectra in Mn12 oriented powders measured at 160.5MHz for three different temperature. Here H is parallel to the easy-axis.

One important issue in molecular nanomagnets is the knowledge of the local spin configuration corresponding to the collective quantum state described by the total spin i.e S=10 for both Mn12 and Fe8. This issue has been addressed very successfully by NMR and we will summarize the results for both Mn12 and Fe8 in the following.

a) Mn12

Figure 22 shows the external magnetic field dependence of the $^{55}$Mn resonance frequencies of the three signals in the spectrum with the magnetic field applied along the easy-axis in Mn12. With increasing parallel field, P1 shifts to higher frequency while the other two peaks P2 and P3 (pertaining to $Mn^{3+}$ ions) shift to lower frequency. Since the resonance frequency is proportional to the vector sum of the internal field ($H_{int}$) and the external field ($H_{ext}$) i.e. $\omega_R = \omega_N(H_{int}+H_{ext})$, this result indicates that the direction of the internal field at the Mn sites for $Mn^{3+}$ ions is opposite to that for $Mn^{4+}$ ions. Since $H_{int}$ originates mainly from the core-polarization [47], $H_{int}$ is negative and the direction of the internal fields at nuclear sites is opposite to that of the Mn spin moment. Thus one can conclude that spin direction of $Mn^{4+}$ ions is antiparallel to the external field, while that of $Mn^{3+}$ ions is parallel to the external field, corresponding to the standard spin structure of magnetic core of Mn12 cluster (see Fig. 23) [52].

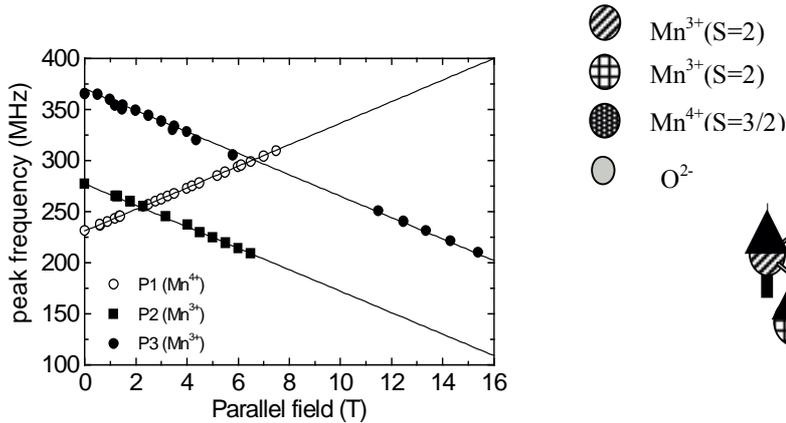
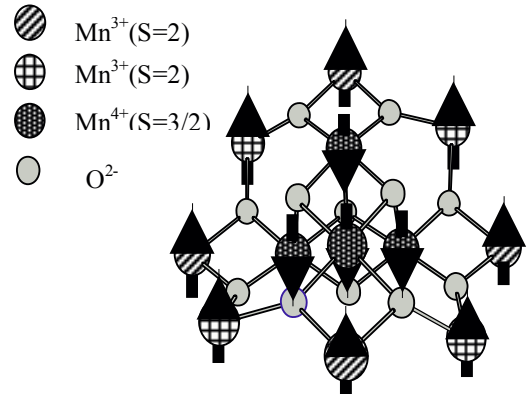

Fig. 22: Parallel field dependence of resonance frequency for each $^{55}$Mn-NMR peak at T=1.5K

Fig. 23: Schematic structure of the magnetic core of Mn12 and orientation of the Mn moments in the ground state.

In the case when the external magnetic field is applied perpendicular to the easy axis (which is the common axis of the oriented powder), the field dependence of the resonance frequencies is the one shown in Fig. 24. As described above, the resonance frequency is proportional to the effective internal field at the nuclear site, which is the vector sum of $H_{int}$ due to spin moments and $H_{ext}$ due to the external field i.e. $|H_{eff}|=|H_{int} + H_{ext}|$. Thus the opposite field dependence of $|H_{eff}|$ for $Mn^{4+}$ and $Mn^{3+}$ ions indicates that the direction of $Mn^{4+}$ spin moments is antiparallel to that of $Mn^{3+}$ spin moments. This leads to the conclusion that the individual spin moments of both $Mn^{4+}$ and $Mn^{3+}$ ions do not cant independently along the direction of the transverse field but rather rotate rigidly maintaining the same relative spin configuration. For a detailed quantitative analysis of the results we refer to the original paper [53].

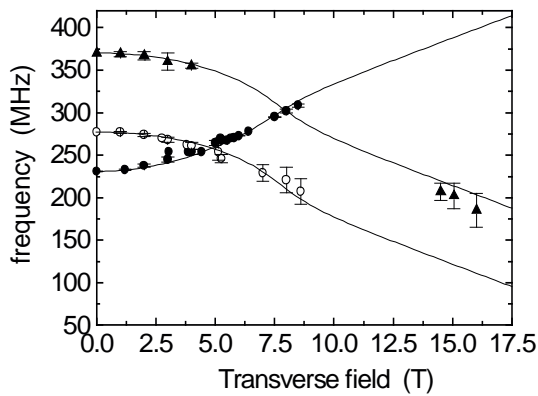

Fig. 24: Transverse field dependence of resonance frequency for each $^{55}$Mn-NMR peak at T=1.5K.

b) Fe8

A similar study has been performed in Fe8 at 1.5K where the nanomagnet is in its ground state. The field dependence of the eight $^{57}$Fe resonance frequencies as a function of a magnetic field applied along the main easy axis is shown in Fig. 25. An analysis analogous to the one performed in Mn12 leads to the conclusion that the internal spin configuration of the ground state is as depicted in Fig. 26 [54]. When the magnetic field is applied perpendicular to the main easy axis and parallel to the medium axis in the xy hard plane the field dependence of the resonance frequencies is quite different as shown in Fig. 27. Again a quantitative analysis of the results leads to the same conclusion as for Mn12 i,e. that the local spin configuration remains intact up to at least 4 Tesla transverse field whereby the effect of the field is the one of rotating rigidly the eight $Fe^{3+}$ moments. For details see ref. 50.

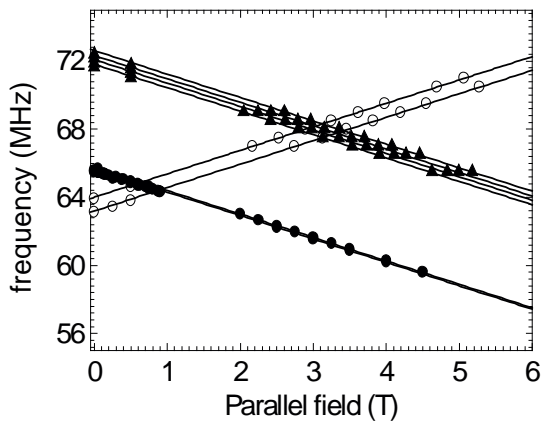

Fig. 25: Parallel field dependence of resonance frequency for each $^{57}$Fe-NMR peak at T=1.5K

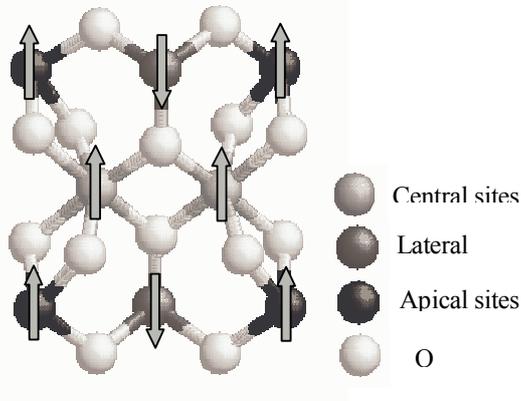 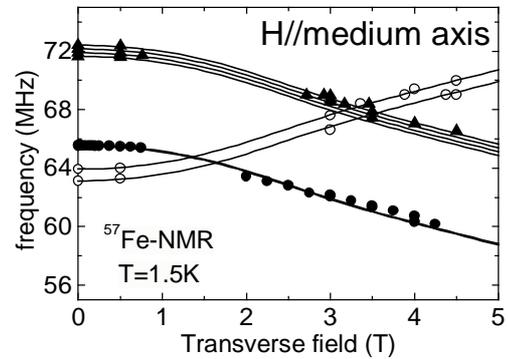

Fig. 26: Structure of Fe8 and orientation of the Fe moments in the ground state.

Fig. 27: Transverse field dependence of resonance frequency for each $^{57}$Fe-NMR peak at T=1.5K.

### c) Spin dynamics and quantum tunneling of the magnetization in nanomagnets: Mn12, Fe8

At very low temperature, when the nanomagnet occupies mostly the magnetic ground state the thermal fluctuations of the magnetization become vanishingly small. In this temperature range the spin dynamics is dominated by quantum fluctuations. In this paragraph we will concentrate on the issue of what information NMR can give about the phenomenon of quantum tunneling of the magnetization (QTM). Let us first briefly summarize the QTM phenomenon. The ground state of both Mn12 and Fe8 clusters is a high total spin ground state i.e. S=10. The S=10 ground state is split into eleven sublevels by a strong easy axis anisotropy [13, 14, 55]. The remaining Kramers degeneracy is removed by an external magnetic field directed along the z easy axis. The energy levels for H // z are obtained from Hamiltonian Eq. 19 as:

$$E_m = -Dm^2 - Bm^4 + g\mu_B H\, m \qquad (26)$$

Assuming g=2 one has $g\mu_B H$ = 1.33H (K) for H in Tesla. For Mn12 one has D=0.55K, B=1.2x10$^{-3}$ K ; for Fe8 one has D= 0.27K, B=0 and in the total Hamiltonian $H = -DS_z^2 - BS_z^4 + g\mu_B \mathbf{H}\cdot\mathbf{S}$, the rhombic term $E(S_x^2 - S_y^2)$, with E=0.046K, must be added. This term modifies the Eq.(26) of the energy levels in a non-simple way.

Below liquid helium temperature the clusters occupy mostly the m = ± 10 states and the reorientation of the magnetization between these two states becomes extremely long – (about one day for Mn12 at 2.4 K) due to the anisotropy barrier giving rise to a pronounced superparamagnetic behavior [13, 14, 55]. When the relaxation rate of the magnetization is measured in response to a varying magnetic field $H_z$ along the easy axis peaks are observed which have been interpreted as a manifestation of resonant tunneling of the magnetization [24, 55]. The qualitative explanation is that the relaxation rate of the magnetization is maximum at zero field and at field values where the total spin states become pairwise degenerate again. The longitudinal field at which this occurs can be easily calculated from Eq. 26 with the parameters given for Mn12 and Fe8 respectively. It is this degeneracy which increases the tunneling probability and thus shortens the relaxation time. The size of the effect depends on terms not shown in the Hamiltonian Eq. 19 , terms which couple the pairwise degenerate states. In particular a transverse magnetic field component can greatly enhance the tunneling splitting of the degenerate levels and thus the QTM. On the other hand the QTM is reduced by the smearing out of the energy levels of the spin states due to spin-phonon coupling, intermolecular interactions and/or hyperfine interactions with the nuclei [56].

NMR can detect the QTM in two quite different ways which will be briefly described in the following.

i) *Measurement of the relaxation rate of the magnetization by monitoring the NMR signal intensity in off-equilibrium state.*

The idea is quite simple. We have seen above that in the NMR spectrum at low temperature in Mn12 and Fe8 the position of the resonance lines depend upon the internal field due to the magnetization of the molecule. Thus if an external field is also applied the position of the line depends on the vector sum of the external field plus the internal field the latter being directed along the magnetization of the molecule. If the direction of the external field is suddenly reversed (or the sample is flipped by 180°) the position of the NMR line changes in the new off-equilibrium situation. The intensity of that particular line starts from zero and it grows back to the full intensity as the magnetization of the molecule relaxes back to equilibrium along the applied field. The method was first described for proton NMR in Mn12 by looking at the echo intensity at the Larmor frequency when the external field is turned on adiabatically at low temperature [57]. A more straightforward implementation was later applied to the signal intensity of shifted proton lines in Mn12 as described below. The detailed results for proton NMR in Mn12 are discussed in ref. 58.

The magnetization of the Mn12 clusters is initially prepared in equilibrium conditions with the magnetic field along the easy c-crystal axis. By inverting the magnetic field or better by flipping the oriented powder by 180° one creates an off-equilibrium conditions whereby the magnetization of each molecule wants to realign along the external field (m=-10 to m=+10 transition). At low temperature and in magnetic fields less than 1 Tesla this process is prevented by the crystal field anisotropy and proceeds very slowly via spin tunneling and phonon assisted relaxation [55,56]. Figure 28 shows the experimental results. The spectrum at the bottom of Fig. 28 corresponds to the thermal equilibrium state before the inversion, where the easy axis of the clusters is along the magnetic field. Just after the inversion of direction of the sample, the observed spectrum changes drastically as shown in the 2nd spectrum from the bottom of Fig. 28. In the figure, the time evolution is from the bottom up. Since the spectra were obtained by sweeping the magnetic field, a process, which takes about 30 minutes, each spectrum, does not correspond to a precise off equilibrium state. However, since the overall process of relaxation of the magnetization at this temperature takes a two or three hundred minutes the different spectra give a qualitative idea of the evolution of the NMR spectrum in time. The signals of the shifted peaks with positive hyperfine fields disappear, while new signals can be observed at magnetic fields higher than the Lamor field $H_0$, where no signal could be detected before the inversion. After a long time (for example, 400 minutes for this case), the spectrum becomes independent of time and it recovers the initial shape before the field inversion.

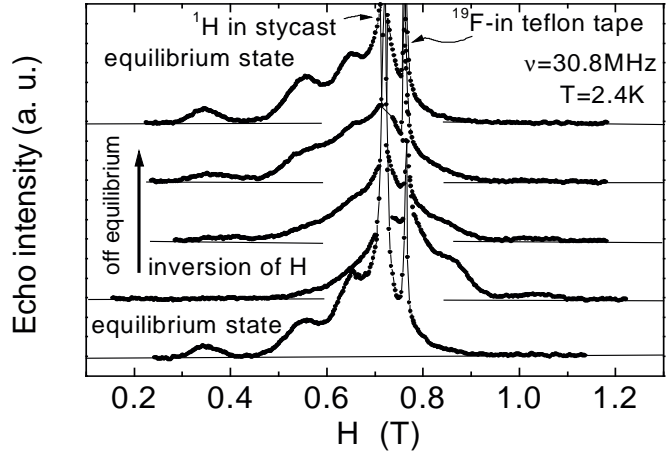

Fig. 28: Time evolution of $^1$H-NMR spectrum of Mn12 oriented powders measured at 30.801MHz and T=2.4K, for H parallel to the easy-axis. The spectrum at the bottom corresponds to the thermal equilibrium state. The second lowest spectrum corresponds to the off-equilibrium situation following the field inversion. The remaining spectra from bottom up are taken at different times after the field inversion.

In order to investigate the effect quantitatively one can sit at fixed field on one of the shifted line (see Fig. 21) and follow its amplitude as a function of time without need of recording the full spectrum. The signal intensity for each shifted peak (P2~P5 shown in Fig. 21) in the spectrum at thermal equilibrium corresponds to the total number of clusters occupying the magnetic m=-10 ground state. Immediately after the $180^0$ rotation of the sample the state m=-10 becomes m=+10. Then the growth of the signal intensity for each peak after the inversion is proportional to the increase of the number of cluster which return to the m=-10 new ground state. Therefore, we can measure a relaxation time of magnetization by monitoring the echo intensity as a function of time. Fig. 29 shows a typical time dependence of the echo intensity h(t) measured at 0.4212T (at the position of the P2 peak) and at T=2.4K. The experimental results can be fitted tentatively by the expression

$$h(t) = a\,(1-\exp\{-t/\tau\,(H)\}) + b \qquad (27)$$

where $\tau(H)$ is a relaxation time and a+b is the echo intensity for the thermal equilibrium state, $h_{T.E.}$. As can be seen in the figure, the growth of the signal intensity is well fitted by a single exponential function (eq. 1) except for a initial fast growth, which accounts for about 30% of the signal. From the slope of the 1- $h(t)/h_{T.E.}$ on semi log plot, we can estimate $\tau(H)$. In the Fig. 30 it is shown the comparison of the field dependence of the relaxation time measured with NMR and the one measured directly for monitoring the magnetization with a SQUID. Although the two sets of data refer to two different ways of extracting the relaxation time from the recovery curves, one can conclude that the results from both methods are in good agreement. In particular, in both cases one sees minima of $\tau(H)$ at the level crossing fields i.e. H=0 (only for the magnetization), H=0.45 Tesla and H=0.9 Tesla. Except for the minima, indicated by the arrows in Fig. 30, the H-dependence of $\tau(H)$ follows a thermal activated law $\tau(H)= \tau_0 \exp\{(67-13.3H)/k_B T\}$ with $\tau_0 \sim 10^{-6}$ (sec) which is consistent with a background thermal excitations over the barrier due to the anisotropy as modified by the applied magnetic field. The field values corresponding to the minima in the relaxation time agree with critical fields where the magnetic level crossing occurs. It should be pointed out that the NMR method to monitor the relaxation of the magnetization of the molecule is not entirely equivalent to the thermodynamic method since the NMR signal is a local probe of the magnetization. This subtle difference suggests that in the NMR method there are microscopic information about the mechanism of reversal of the magnetization. The method has been confirmed [48, 59, 60] by measuring the $^{55}$Mn NMR spectrum in Mn12 with results very similar to the ones obtained with proton NMR [58] and it has been extended to very low temperature to observe the avalanche effect of the spin reversal in the magnetization recovery in Mn12 [61]. Also it has been applied to proton NMR in Fe8 at very low temperature to obtain information about Landau-Zener transition as the field is swept through a level crossing condition [62].

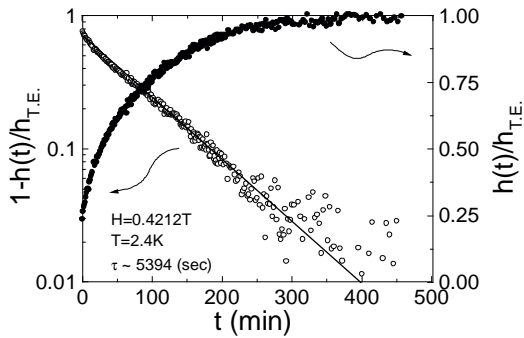
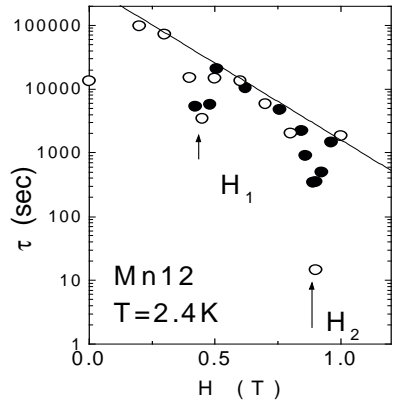

Fig. 29: Time dependence of the echo intensity in Mn12 oriented powders taken at H=0.4212T and T=2.4K following the sudden inversion of the orientation of the easy axis of the sample with respect to the applied magnetic field. Solid and open circles are $h(t)/h_{T.E.}$ and 1- $h(t)/h_{T.E.}$, respectively.

Fig. 30: Field dependence of the relaxation time for Mn12 (oriented powders) magnetization measured at T=2.4K by $^1$H-NMR (closed circles) and SQUID (open circles). The solid line is $\tau(H)= \tau_0 \exp\{(67-13.3H)/k_B T\}$ discussed in the text. H is parallel to the easy-axis.

*ii) Nuclear spin-lattice relaxation induced by quantum tunneling of the magnetization in molecular nanomagnets : Fe8, Mn12*

Quantum fluctuations of the magnetization are expected to have a distinctive effect on nuclear spin-lattice relaxation (NSLR). For example in the case of Fe8 the proton NSLR was observe to decrease very fast down to 400mK as the result of slowing down of the thermal fluctuations but then to become temperature independent below 300mK [62]. This leveling off of proton $T_1$ was taken as indication of a crossover to a regime of quantum fluctuations of the magnetization. A more direct way to detect the effect of quantum tunneling on NSLR is obtained from the field dependence of $T_1$ as illustrated in the following.

The occurrence of QTM in zero external magnetic field or in an external field applied along the anisotropy axis is related to the splitting of the pairwise degenerate magnetic levels by an amount $\Delta_T$ that is due to off diagonal terms in the magnetic Hamiltonian arising from anisotropy in the xy plane, intermolecular dipolar interactions and hyperfine interactions. Normally the tunnel splitting $\Delta_T$ in the Mn12 and Fe8 clusters is much smaller than the level broadening so that measurements of $\Delta_T$ is difficult. The effect of QTM as a function of the longitudinal magnetic field can be seen as dips in the relaxation time of the magnetization in correspondence of the critical fields for level crossing as shown in Fig. 30. However, under applied parallel field no clear effect could be observed in the proton $T_1$. This can be seen in Fig. 31 where we show the longitudinal field dependence of $T_1$ in both Mn12 and Fe8. The critical fields corresponding to the first level crossing in the ground state (i.e., *m*=-10 to *m*=9) is estimated to be =0.5T and 0.2T for Mn12 and Fe8 respectively from Eq. 23. As seen in Fig. 31 the data can be fit well by the simple model of thermal fluctuations of the magnetization described in Paragraph IV b without any marked anomaly at the level crossing fields. Thus we may conclude tentatively that for **H//z** no effects of quantum tunneling can be observed on $1/T_1$ since the tunneling dynamics is too slow for longitudinal applied fields. As a word of caution we may add that the data in Fig. 31 indicate some enhancement of the NSLR in Mn12 around the level crossing field and in Fe8 the data are incomplete since the first level crossing cannot be reached as the result of the experimental difficulties at low fields thus leaving open the possibility that a small effect of QTM may be observed for longitudinal fields. On the other hand a recent study of $^{55}$Mn NMR in Mn12 at very low temperature ( down to 20 mK) has revealed the presence of ( temperature independent ) tunneling fluctuations [62]

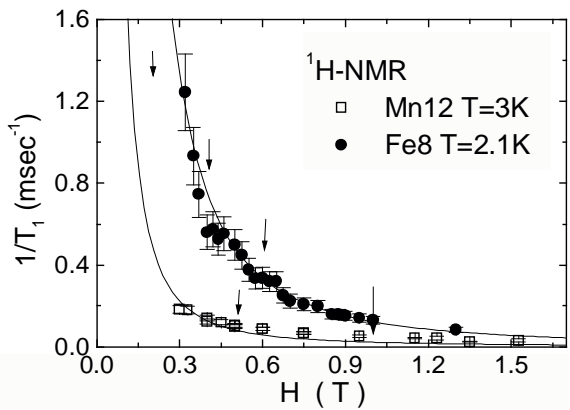

Fig. 31: Parallel field dependence of $^1$H-$1/T_1$ in both Mn12 (T=3K) and Fe8 (T=2.1K), oriented powders. The arrows indicate the critical field values for level crossing across the energy barrier

On the other hand by applying a magnetic field perpendicular to the easy axis (transverse field), one can increase $\Delta_T$ of all levels while leaving the symmetry of the double well potential intact [56,63]. An increase of the tunneling splitting corresponds to an increase of the tunneling frequency or of the tunneling probability. For H>1T, in Fe8, the relaxation (fluctuation) of the magnetization driven by tunneling (coherent and/or incoherent) becomes so fast that it falls within the characteristic frequency domain (MHz) of a NMR experiment. Therefore when the magnetic field is applied perpendicular to the main easy axis z (transverse field) a pronounced peak in the proton spin-lattice relaxation rate, $1/T_1$, of protons in a single crystal of Fe8 as a function of external magnetic field can be observed at 1.5K as shown in Fig. 32. The effect is well explained by considering that by increasing the transverse field the incoherent tunneling probability becomes sufficiently high as to match the proton Larmor frequency. When the applied field goes through this condition the fluctuation rate of the magnetization is most effective in driving the nuclear relaxation and a maximum appears in $1/T_1$. The peak disappears when a parallel field component is introduced in addition to the transverse field, by tilting the single crystal about 5 degrees in yz plane. (see Fig. 32). Since the parallel field component removes the degeneracy of the ±$m$ magnetic states and consequently the possibility of tunneling it is clear that the peak of $1/T_1$ must be related to a contribution to the nuclear relaxation rate from the tunneling dynamics. A quantitative interpretation of the effect and further details can be found in ref. 64.

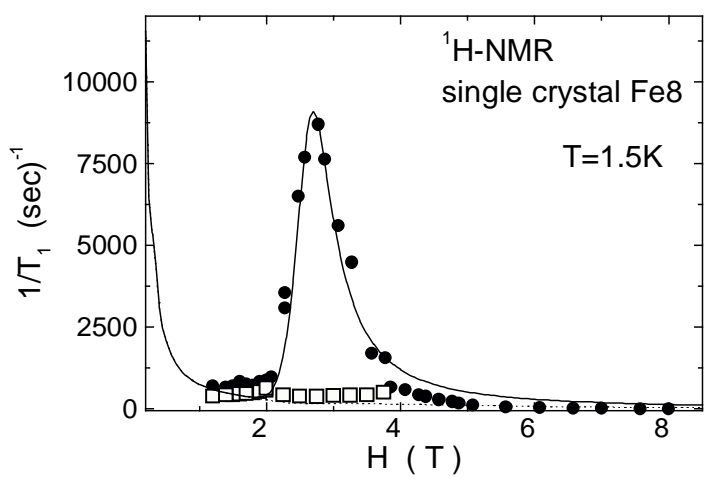

Fig. 32: Transverse field dependence of $^1$H-$1/T_1$ at T=1.5K in Fe8 single crystal as a function of the field along the medium axis (closed circles). Open squares are the results obtained when the single crystal is tilted so that the applied field is 5 degrees off the xy-plane.

Measurements of NSLR as a function of transverse field were also performed in oriented powder of Mn12 both on proton NMR and on $^{55}$Mn NMR [53]. However, in this case we failed to observe a peak of the NSLR as for Fe8 although the relaxation rate was found to be faster than expected on the basis of purely thermal fluctuations of the magnetization. The results are shown in Fig. 33 where we plot, the transverse magnetic field dependence of $^{55}$Mn-$1/T_1$ (P1; $Mn^{4+}$ ions). With increasing the transverse magnetic field, $1/T_1$ increases rapidly about two decades and shows a broad maximum around 5.5T. We have calculated the expected field dependence obtained from the simple thermal fluctuation model described in Paragraph IV b a model which works well to describe the longitudinal field dependence. As seen from the comparison of the dotted line and the experimental points in Fig. 33, there appears to be an additional contribution to NSLR of $^{55}$Mn in transverse field. The additional contribution was explained [53] by a phenomenological model which considers the effect of canting of the magnetization in presence of a transverse field. In fact the quantization axis of the nuclear spin does not coincide with that of the internal magnetic field due to the electron spins. Therefore the transverse field generates components of the local field perpendicular to quantization axis of the nuclear spin, whose fluctuation can be very effective in producing nuclear relaxation. For the details of the model which generates the fitting curve in the figure we refer to ref. 53.

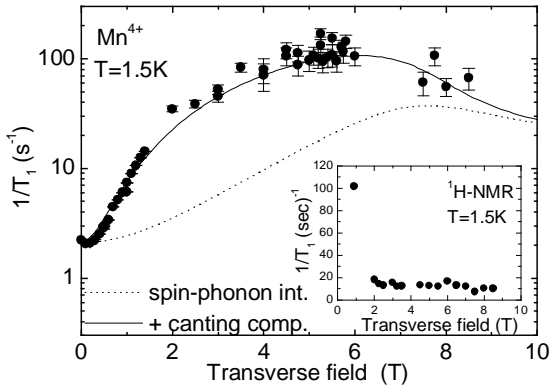

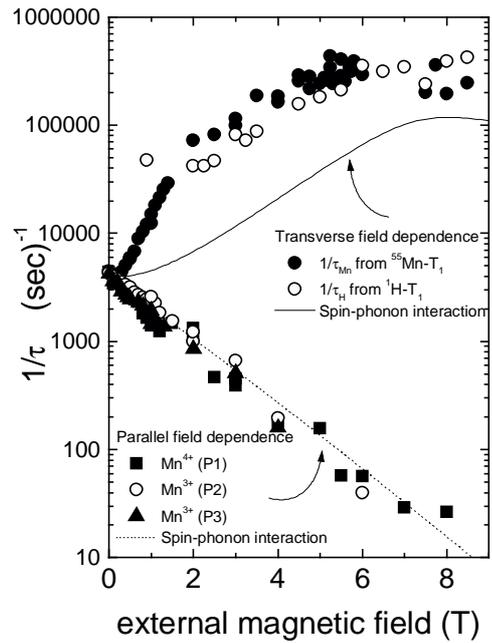

Fig. 33: Transverse field dependence of $1/T_1$ for $^{55}$Mn-NMR at the $Mn^{4+}$ site (P1) measured at T=1.5K in Mn12. The inset shows the transverse field dependence of $1/T_1$ of $^1$H-NMR.

Fig. 34: External magnetic field dependence of $1/\tau$ in Mn12 oriented powders, estimated from a relation of $T_1^{-1}=A/(\omega_N^2\tau)$ for all case. Solid and dotted lines are transverse and parallel field dependence of $\tau^{-1}$ calculated from the spin phonon interaction.

The transverse field dependence of $1/T_1$ for protons at 1.5K in Mn12 is also shown in the inset of Fig. 33. Contrary to $^{55}$Mn, the proton $1/T_1$ has a rapid initial decrease with increasing field, roughly as $H^{-2}$, which can be explained very well by Eq. 18 in the limit of "slow motion" ($\omega_N \gg 1$), where $T_1^{-1}$ can be approximated with $A/(\omega_N^2)$. The key to understand the different behavior of $1/T_1$ for $^{55}$Mn and for $^1$H is the fact that the internal magnetic fields at proton sites are very small (at most 0.4 T in comparison with the case of Mn nuclei, ~22T for $Mn^{4+}$ ions) and thus $\omega_N \propto H_{ext}$ for protons while $\omega_N$ depends very little from the external field for manganese. At higher fields the proton NSLR levels off to an almost constant value and no peak can be observed contrary to Fe8 (see inset of Fig. 33). The peak due to the matching of the Larmor frequency with the tunneling probability was expected in Mn12 in transverse field around 6-7 T. We do not have as yet an explanation for the different behavior of Mn12 and Fe8 except for the fact that the data in Mn12 were taken in oriented powder since no single crystal large enough for NMR work is available for Mn12. The leveling off of $1/T_1$ for protons at high fields is due to the same decrease of the lifetime $\tau$ which is the dominant effect for $^{55}$Mn. This can be seen by extracting $\tau$ from the equation of $T_1^{-1}=A/(\omega_N^2)$ and the experimental points for both $^{55}$Mn and $^1$H. The two sets of data coincide within experimental error as shown in Fig. 34. In the figure, we show also for comparison the transverse field dependence of $\tau^{-1}$ calculated from the spin phonon interaction. The difference of the experimental points from the solid line represent the contribution to the fluctuation of the magnetization which cannot be related to spin-phonon interaction.

## VI. Miscellaneous NMR studies of molecular clusters: Fe2, Fe4, Fe30, Ferritin core, Cr4, Cu6, V6, V15.

In this paragraph we will briefly review the NMR work done in a number of magnetic molecular clusters that were not included in the previous paragraphs because they present specific features which do not fit entirely into the general issues discussed above.

**i) Iron clusters : Fe2, Fe4, Fe30 and Ferritin core**
The Iron(III) S=5/2 dimer, [Fe(OMe)(dbm)$_2$]$_2$ (in short Fe2), has a nonmagnetic S=0 ground state. The separation between the singlet ground state and the first excited (triplet) state was determined by susceptibility measurements to be about 22K; proton NMR measurements were performed on Fe2 [65]. The nuclear spin-lattice relaxation rate (NSLR) was studied as a function of temperature at 31 and 67 MHz and as a function of the resonance frequency (10-67MHz) at T=295K. The results are shown in Fig. 35. At room temperature the $^1$H NSLR is independent of frequency (see inset in Fig. 35) contrary to the strong field dependence found in AFM rings as discussed in Paragraph III. The temperature dependence of the proton NSLR shows a monotonous decrease on lowering the temperature without the peak characteristic of the AFM as described in Paragraph IV a. $1/T_1$ is approximately proportional to $\chi T$ where $\chi$ is the uniform susceptibility measured at H=1T.

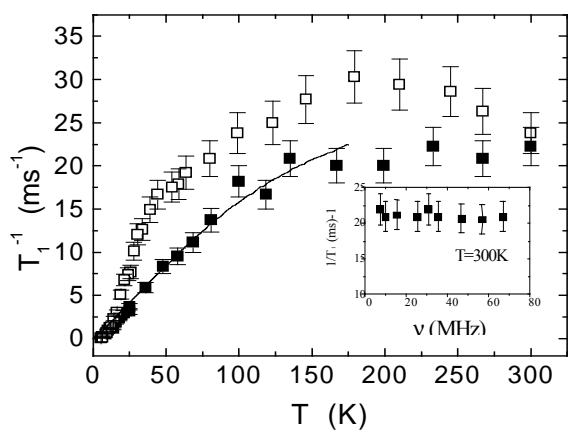
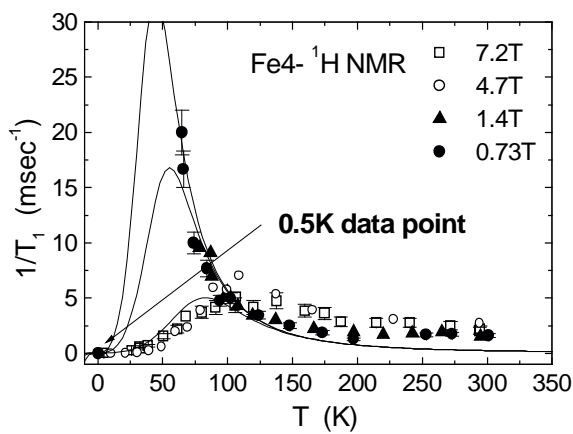

Fig. 35: Temperature dependence of $^1$H-$1/T_1$ in Fe2 powders. The inset shows magnetic field dependence of $1/T_1$ at room temperature.

Fig. 36: Temperature dependence of $^1$H-$1/T_1$ in Fe4 "oriented" powders.

The cluster of four Iron(III) ions, $Fe_4(OCH_3)_6(dpm)_6$ (in short Fe4), is characterized by a total spin ground state $S_T = 5$ and Ising anisotropy [66]. The cluster behaves at low temperature like a superparamagnet just like Mn12 and Fe8 except that the anisotropy barrier is much lower. The D=0.29K in Eq. 26 corresponds to an energy barrier between the lowest $M_S = \pm 5$ state and the highest $M_S = 0$ state of only 7.25 K. The Fe4 cluster has been investigated by $^1$H NMR as a function of temperature (0.5 - 295K) and external magnetic field (0.3 - 7.2T) [67]. The results of NSLR are shown in Fig. 36. At very low T (0.5K) the spectrum becomes very broad indicating the freezing of the $Fe^{3+}$ moments in a superparamagnetic state. The temperature dependence of $T_1^{-1}$ is characterized by a field dependent maximum as shown in Fig. 36. For low magnetic fields the peak of $1/T_1$ becomes so high that the proton NMR signal cannot be detected in the temperature region of the peak. The field dependent maximum as well as the loss of NMR signal the region of the peak at low magnetic fields (see Fig. 36) is qualitatively similar to the one observed in Mn12 [68]. In Mn12, $^{13}$C [69] and $^2$D NMR [70, 71] were also performed giving similar qualitative experimental results. In ref. 71 an isotope effect on NMR data was also suggested. It should be noted that the field dependent peaks observed in Fe4 and in Mn12 with maxima located at temperatures of the order of the exchange constant J have many similarities with the peaks discussed for AFM rings in Section IV. A systematic analysis of the nuclear relaxation data in ferrimagnetic molecular clusters is in order to establish the relaxation mechanism at these intermediate temperatures.

The cluster $\{Mo_{72}Fe_{30}O_{252}(Mo_2O_7(H_2O))_2(Mo_2O_8H_2(H_2O))\text{-}(CH_3COO)_{12}(H_2O)_{91}\}\cdot$ $\cdot 150H_2O$ [2], Fe30 in brief, has 30 Fe(III) ions occupying the 30 vertices of an icosidodecahedron. The magnetic properties are characterized by a ground state with total spin state $S_T = 0$ due to O-Mo-O bridges mediating antiferromagnetic (AF) coupling with the exchange coupling constant J = 1.57 K between nearest-neighbor Fe ions. An accurate description of the magnetic properties of the cluster has been based on classical Heisenberg model of spins on the vertices of an icosidodecahedron for arbitrary magnetic fields [2,72]. $^1$H nuclear magnetic resonance (NMR) and relaxation measurements have been performed in the Keplerate species Mo72Fe30 [73]. The $^1$H NMR linewidth increases gradually with decreasing T and it saturates below about 4 K, as expected for a non-magnetic ground state with $S_T = 0$. The results for the magnetic field and temperature dependence of $T_1^{-1}$ are shown in Figs. 37(a) and (b). The magnetic field dependence of $T_1^{-1}$ at room temperature (see Fig 37(a)) can be fitted by Eq. 7 used to fit the results in AFM rings. The best fit constants A= 5 ms$^{-1}$, B = 0.1 Tesla and C= 2.2 ms$^{-1}$ compare well with the constants in Table I. The small value of B, corresponding to $\Gamma_A = \gamma_e B = 1.7\ 10^{10}$ (rad/sec), is consistent with a highly isotropic Heisemberg system. As shown in Fig. 37(b) a strong enhancement of $T_1^{-1}$ with a peak around T = 2 K is observed, a feature similar to the one observed in antiferromagnetically (AFM) coupled rings (see Paragraph IV a). It must be mentioned that the first direct experimental confirmation of the transition from $S_T=0$ to $S_T=1$ for standard Fe30, was from µSR experiments where, in correspondence to the first crossing field $H_c \sim 0.24$ Tesla, the width of the distribution of the local fields at the muon site tends to a plateau [74].

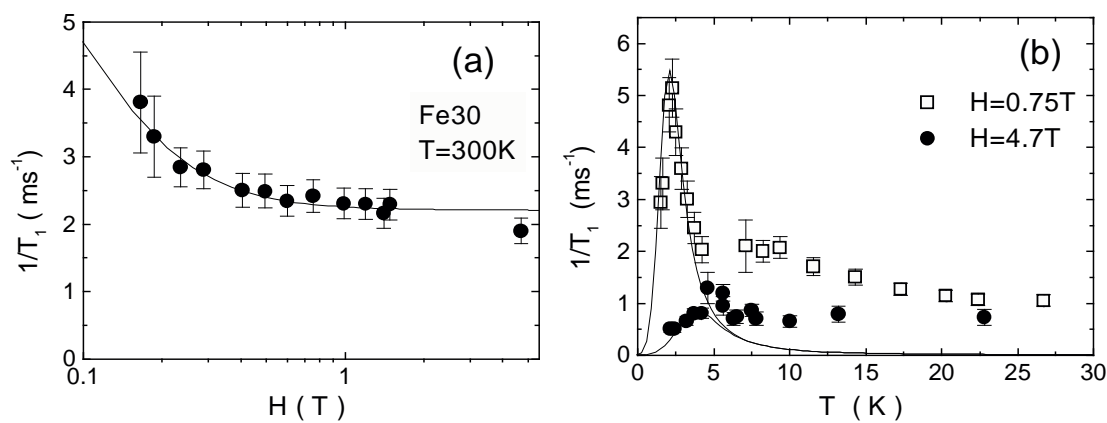

Fig. 37: (a) Magnetic field dependence of $^1$H-1/$T_1$ at 300K in Fe30 powders. (b) Temperature dependence of 1/$T_1$ measured at H=0.75K (open squares) and H=4.7T (closed circles).

An antiferromagnetic molecular cluster containing a large number of Fe(III) ions (a variable number of the order of 5000) is the biomolecule ferritin. It is commonly reported [75] that ferritin becomes antiferromagnetic (AF) although the evidence is largely indirect and the Neel temperature reported varies widely in the range 50K < $T_N$ < 240K. Moreover, for T < 30K, magnetization measurements [76] give evidence of superparamagnetic relaxation and of spin freezing of the total magnetic moment associated with the uncompensated spins at the boundary of the cluster of iron ions. Recently the interest in the magnetic behavior of ferritin has been revived by the confirmation [77, 78] of an early report [79] of the occurrence of macroscopic quantum coherence phenomena at low temperature. Since the naturally occuring ferritin biomolecule contains the largest number of protons in the outer protein shell and thus are not coupled to the iron ions in the core, one has to strip the protein shell in order to use the NMR of the protons inside the core to probe the magnetic behavior of the $Fe^{3+}$ ions.

Proton NMR were performed in ferritin core as a function of temperature from 220K down to 4.2K at 4.7 T [80]. The proton spectrum is inhomogeneously broadened and the NSLR depends on the position in the spectrum where the measurements are performed making the analysis of the results difficult and uncertain. The relevant result of the investigation is represented by the maximum in the NSLR observed at a temperature of about 100K which is a clear indication of the occurrence of an antiferromagnetic transition. The study has to be viewed as preliminary. More experiments are needed to investigate the critical behavior of the NSLR around the ordering temperature and the effect of the superparamagnetic fluctuations and quantum tunneling.

**ii) Ferromagnetic (FM) ring (Cu6) and cluster (Cr4).**

Both $[(PhSiO_2)_6Cu_6(O_2SiPh)_6]$ (in short Cu6) and $[Cr_4S(O_2CCH_3)_8(H_2O)_4](NO_3)_2 \cdot H_2O$ (in short Cr4) are molecular nanomagnets with nearest neighbor ferromagnetic exchange interaction. Cu6 is a planar ring of six spins ½ $Cu^{2+}$ ions with J=61K and a ground state with $S_T = 3$ separated by an energy gap of 30K from the first excited state $S_T = 2$ [81,82]. The ground state is split by an anisotropy D=0.435 K. Thus in presence of an external magnetic field H, the energy levels of the 7 magnetic substates m in $S_T = 3$ state are given by:

$$E_m/k_B = 0.435 \cdot m^2 - 1.33 \cdot m \cdot H \quad \text{(Kelvin, with B in Tesla)}. \tag{28}$$

On the other hand Cr4 is a cluster formed by four spin 3/2 $Cr^{3+}$ ions on the vertices of a almost regular tetrahetron [83]. The exchange constant J=28K and the ground state has $S_T = 6$. The crystal field anisotropy is negligibly small.

Measurements of proton NMR have been performed in both systems in the low temperature range where the magnetic clusters are in their total $S_T$ ground state [82, 83]. It is noted that contrary to Mn12 and Fe8 no superparamagnetic effects are observable since the anisotropy for Cu6 is positive and for Cr4 is negligible. Thus no energy barrier is present for the reorientation of the magnetization even at low temperature. These FM cluster could be thought as soft nanomagnets in contrast to Mn12 and Fe8 which are hard nanomagnets. Therefore in the temperature range of our measurements (1.4-4.2K) the thermal fluctuations of the magnetization is still faster than the NMR frequency and no static local field is present at the proton site (i.e., no spin freezing). The systems behave like a assembly of very weakly interacting paramagnetic spins of size $S_T=3$ and $S_T=6$ for Cu6 and Cr4 respectively. The results for the proton NSLR are shown in Fig. 38(a) and Fig. 39(a) for Cu6 and Cr4 respectively.

The NSLR results can be explained very well by a simple model of spin-lattice relaxation due to the fluctuations of the total magnetization of the cluster. One can assume that the transverse hyperfine field at the proton site due to the interaction with the different Cu (Cr) magnetic moments is proportional to the component of the magnetization of the cluster in the direction z of the applied field i.e. $h_\pm(t) = C \cdot M_z(t) = C \cdot (\langle M_z(t)\rangle + \delta M_z(t))$. Then from Eq. 14 it follows for the NSLR:

$$1/T_1 = (1/2)\gamma^2 C^2 \int \langle\delta M(0)\rangle^2 \exp(-\Gamma t)\exp(-i\omega_L t)dt = A\langle\delta M(0)^2\rangle \Gamma/(\Gamma^2 + \omega_L^2) \quad (29)$$

where we dropped the subscript z for simplicity. The characteristic frequency $\Gamma$ can be viewed classically as a probability for the total spin $S_T = 3$ (6) of the Cu6 (Cr4) cluster to change its orientation along the external magnetic field or, quantum mechanically, as the broadening of the corresponding magnetic eigenstate. The average of the square of the fluctuation can be calculated exactly as:

$$\langle\delta M(0)^2\rangle = \frac{k_B T}{(g\mu_B)^2}\chi(H,T) = \sum_m \frac{m^2 \exp(-E_m/kT)}{Z} - \left(\sum_m \frac{m \cdot \exp(-E_m/kT)}{Z}\right)^2 \quad (30)$$

where $E_m$ is given by Eq. 28. The data in Figs. 38(a) and 39(a) have been replotted in Figs. 38(b) and 39(b) as a function of $g\mu_B H/k_B T$. In Cr4 the NSLR data at different fields and temperatures rescale well with $g\mu_B H/k_B T$ indicating that $\Gamma \gg \omega_L$ in Eq. 29. The data are reproduced very well by a theoretical curve obtained from Eq. 29 and Eq. 30 assuming $\Gamma$ to be H and T independent and using the energy levels, Eq. 28, with negligible anisotropy. On the other hand the data in Cu6 shown in Fig. 38(b) do not scale so well vs $g\mu_B H/k_B T$. This is attributed to the non negligible anisotropy term in Eq. 25 and the B and T dependence of $\Gamma$ [82]. In conclusion the above results of proton NSLR demonstrate that in the FM molecular clusters Cu6 and Cr4 the low temperature spin dynamics can be described as a simple thermal fluctuation of the magnetization of the cluster in its collective quantum state of total spin $S_T$. The simple model for the NSLR leads to the same result as obtained from the first-principles calculation of the equilibrium two-ion time correlation function for the given set of magnetic eigenstates of the molecule whereby the characteristic frequency $\Gamma$ for the fluctuations of the total spin of the molecule corresponds to the broadening of the magnetic eigenstates [83].

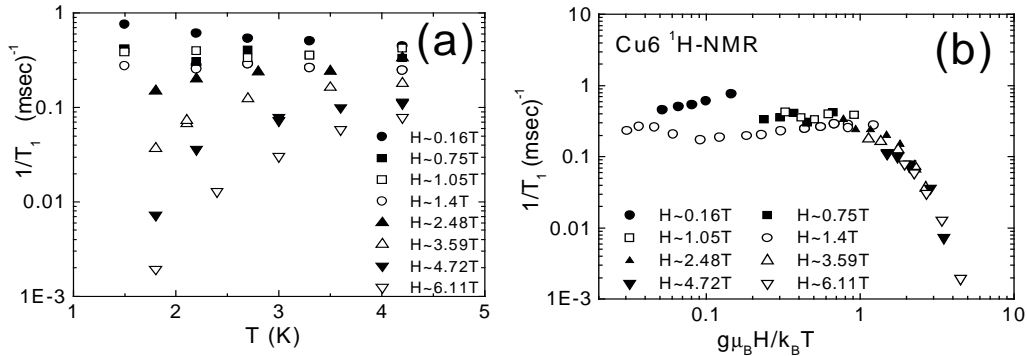

Fig. 38: $^1$H-$1/T_1$ in Cu6 powders. (a) $1/T_1$ plotted as a function of temperature for assorted values of the magnetic field. (b) $1/T_1$ plotted as a function of $g\mu_B H/k_B T$.

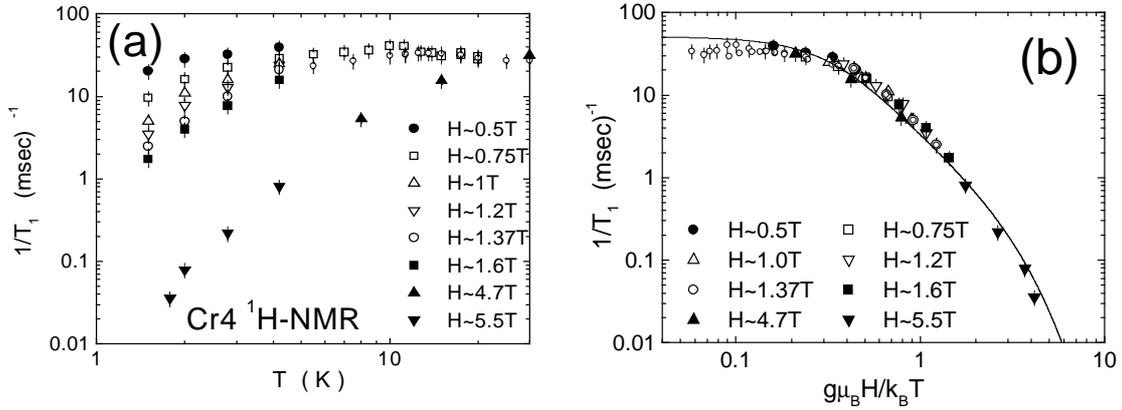

Fig. 39: $^1$H-$1/T_1$ in Cr4 powders. (a) $1/T_1$ plotted as a function of temperature for assorted values of the magnetic field. (b) $1/T_1$ plotted as a function of $g\mu_B H/k_B T$.

### iii) Vanadium clusters: V6 and V15

V15 cluster (complete formula : $K_6[V_{15}As_6O_{42}(H_2O)] \cdot 8H_2O$) contains 15 paramagnetic vanadium ions $V^{4+}$, each of which has spin s= 1/2. The V15 cluster is a very peculiar Heisenberg spin-triangles system which can be viewed as a triangle formed by three groups of five $V^{4+}$ ions [1]. As a result of frustration the ground state is formed by two $S_T$ = 1/2 doubly degenerate states separated by a small gap of about 0.2 K. One expects that the magnetic properties and the spin correlations are dominated by interlayer coupling rather than by intralayer coupling. This situation can be compared with another spin-triangles system namely V6 [84]. In V6 the ions are arranged in two almost independent trinuclear units each forming a strongly antiferromagnetic frustrated triangle. The low temperature ground state for each of these triangles can be characterized by a doubly degenerate $S_T$ = 1/2, a doubly degenerate first excited state with $S_T$ = 1/2 separated by a gap of about 60 K, and a fourfold degenerate second excited state with $S_T$ = 3/2 separated by 85 K [84]. $^1$H NMR and relaxation measurements were performed in both clusters [85, 86], in form of powders. For V6 the T-dependence of proton $T_1^{-1}$ is approximately proportional to $\chi T$, where $\chi$ is the uniform susceptibility. On the other hand, in V15 the T-dependence of $T_1^{-1}$ is quite different from the one of $\chi T$. This was taken as indication of a critical T-dependence of $\Gamma$, a parameter which measures the width of the $V^{4+}$ spin fluctuation spectral density. For details in these two frustrated spin triangles we refer to the refs [85, 86].

A proton NMR study has been performed also in a frustrated triangular system of Mn spins $[Mn_3O(O_2CCH_3)_6(C_5H_5N)_3] \cdot C_5H_5N$, in brief Mn3 [87], powders. The Mn system is similar in many respect to the frustrated spin-triangle system V6 although the three Mn ions in the triangle are different with one $Mn^{2+}$ (s=5/2) and two $Mn^{3+}$ (s=2) ions. Also the coupling constant J' between the two $Mn^{3+}$ spins is three times as large as the coupling J between the $Mn^{3+}$ and the $Mn^{2+}$ spins. The most remarkable NMR result is a strong enhancement of $1/T_1$ with a peak at $T_0 \approx J$ just like in the AFM rings (see Paragraph IV a).

Another frustrated spin-triangle system is $(NH_4)[Fe_3(\mu_3\text{-}OH)(H_2L)_3(HL)_3]$ (in short Fe3, L=neutral ligand) [88]. Here again one observes a peak in $T_1^{-1}$ at T~J, analogously to Mn3.

## VII. Summary and conclusions

In the present review we tried to illustrate in some details a large body of NMR data obtained by our groups starting back in 1996 in magnetic molecular clusters ( or molecular nanomagnets or single molecule magnets). We tried to quote briefly also work done by other NMR groups although some reports may have escaped our attention and we apologize for this. As a conclusion of the review we may say that the main characteristics of NMR in these new interesting magnetic systems have been established and the field is now mature for a deeper investigation and a better understanding of some remarkable effects reported here. In fact the theoretical analysis of the results in this review has been often rather qualitative and based on simple models to describe the NMR and relaxation effects observed. We will summarize in the following some of the major issues emerging from our investigation with the focus on what remains to be done .

We have observed a magnetic field dependence at room temperature of proton $T_1$ in all nanomagnets investigated. Since at room temperature molecular clusters have to be viewed as nanosize paramagnets the field dependence is a clear signature of the zero dimensionality of the system. From the field dependence of $T_1$ one can obtain the "cut-off" frequency for the time decay of the electron spin correlation function. The detailed origin of the "cut-off" frequency for the different molecules remains to be established.

At intermediate temperatures namely when the electron spin in the molecular cluster becomes strongly correlated ( $k_BT \approx J$ ) we have observed a T and H behavior of $T_1^{-1}$ which can be explained satisfactorily with simple phenomenological models based on direct relaxation process. The models are based on the assumption that the relaxation process is a direct process due to the hyperfine field fluctuations related to the lifetime of the magnetic molecular states. In particular for the cases of Mn12 and Fe8 the low temperature nuclear relaxation can be described in terms of fluctuations of the magnetization in the total spin S=10 ground state manifold due to spin-phonon interaction, and a value for the spin-phonon coupling constant could be derived for these two nanomagnets. The fact that the nuclear $T_1$ can be explained in such a simple way came as a surprise since normally in highly correlated spin system such as ferro and antiferromagnetc bulk system close to the phase transition one has to consider the interplay of fluctuations of different symmetry i.e different q-vectors. In nanosize molecular magnets it appears that the discreteness of the magnetic energy levels plays an important role whereby only the lifetime of the total spin quantum state of the molecule is all it matters in determining the nuclear spin- lattice relaxation. One issue which remains to be clarified is the possibility of introducing a spin-wave description of the excited states in these nanomagnets and consequently the relation between a description of the relaxation in terms of a direct process and the description in terms of one and two magnons scattering .

At even lower temperature ($k_BT \ll J$) we have observed dramatic effects on the nuclear relaxation rate due to level crossing effects and quantum tunneling of the magnetization. It appears that NMR can be useful to distinguish between level crossing (LC) and level anticrossing (LAC) in AFM rings but a general theory to describe the observed enhancement of $T_1^{-1}$ at the critical field is still lacking. The same is true for quantum tunneling of the magnetizatiomn. The phenomenon can be detected by NMR in different ways but a detailed description of the coupling between the nuclear Zeeman reservoir and the tunneluing reservoir has to be firmly established in order to extract information from NMR measurements.

Finally it should be mentioned that the NMR spectra at low temperature give important information about hyperfine interactions. In particular the zero field NMR and the evolution of the spectrum as a funtion of an applied field in sngle crystals or oriented powders is a unique method which unables to determine the size and orientation of the local magnetic moments of the molecular cluster in its ground state configuration.

**Acknowledgements**

We thank all collaborators who made possible this investigation and whose names appear in the many papers quoted in the review. A special thank to our colleagues chemists who introduced us to this field and provided the many samples and helped us to understand their properties. Their names also appear in most of our joint publications.

# REFERENCES

[1] D. Gatteschi et al., Science **265**, 1055 (1994).
[2] A. Muller et al., Chem. Phys. Chem **2**, 517 (2001).
[3] K. Wieghardt, K. Pohl, H. Jibril and G. Huttner, Angew. Chem. Int. Ed. Engl. **23**, 77 (1984).
[4] T.Lis, Acta Cryst. **B36**, 2042 (1980).
[5] D.Gatteschi, A.Caneschi, R.Sessoli and A.Cornia, Chem.Soc.Rev. **2**, 101 (1996) ; K.L. Taft, C.D. Delfs, G.C, Papaefthymoiu, S. Foner, D. Gatteschi and S.J. Lippard, J.Am.Chem.Soc. **116**, 823 (1994).
[6] T. Moriya, Prog. Theor. Phys., **16**, 23 (1956); ibidem 28,371 (1962).
[7] F. Borsa and M. Mali, Phys. Rev. **B 9**, 2215 (1974).
[8] J. H. Luscombe and M. Luban, J. Phys.: Condens. Matter **9**, 6913 (1997).
[9] J. Tang, S. N. Dikshit, and J. R. Norris, J. Chem. Phys. **103**, 2873 (1995).
[10] J.H. Luscombe, M. Luban and F. Borsa, J. Chem. Phys. **108**, 7266 (1998).
[11] J. P. Boucher, M. Ahmed Bakheid, M. Nechschtein, G. Bonera, M. Villa, and F. Borsa, Phys. Rev. **13**, 4098 (1976).
[12] A. Lascialfari, Z.H. Jang, F. Borsa, D. Gatteschi and A. Cornia, J. Appl. Phys. **83**, 6946 (1998).
[13] A.L. Barra, P. Debrunner , D. Gatteschi , C.E. Schulz, R. Sessoli, Europhys.Lett.**35**, 133 (1996).
[14] R. Sessoli, Hui-Lien Tsai, A. R. Schake, Sheyi Wang, J. B. Vincent, K. Folting, D. Gatteschi, G. Christou, and D. N. Hendrickson, J. Am. Chem. Soc. **115**, 1804 (1993).

[15] J. Van Slageren et al., Chem. Eur. J. **8**, 277 (2002) ; S.Carretta , J. van Slageren , T. Guidi , E. Liviotti , C. Mondelli , D. Rovai , A. Cornia , A.L. Dearden , F. Carsughi , M. Affronte , C.D. Frost , R.E.P. Winpenny, D. Gatteschi , G. Amoretti , R. Caciuffo, Phys.Rev. **B67**, 094405 (2003).

[16] A. Lascialfari, D. Gatteschi, A. Cornia, U. Balucani, M.G. Pini and A. Rettori, Phys. Rev. **B 57**, 1115 (1998)

[16b] D.Procissi, A.Shastri, I.Rousochatzkis, M.Al Rifai, P.Kogerler and M.Luban, umpublished.

[17] A. Abragam, "The principles of Nuclear Magnetism", Clarendon Press (Oxford, 1961).

[18] T. Moriya, Prog. Theor. Phys. **28**, 371 (1962).

[18b] F.Borsa and A.Rigamonti, in Magnetic Resonance of Phase Transitions, Academic Press (1979)

[19] A.Lascialfari, D.Gatteschi, F.Borsa and A.Cornia, Phys. Rev. **B 55,** 14341 (1997).

[20] D.Procissi, B.J.Suh, E.Micotti, A.Lascialfari, Y.Furukawa and F.Borsa, J.Magn. Magn.Mater., in press (ICM2003).

[21] S. H. Baek, M. Luban, A.Lascialfari, Y.Furukawa, F. Borsa, J.Van Slageren and A.Cornia , unpublished.

[22] B. Barbara, W. Wernsdorfer, L. C. Sampaio, J. G. Park, C. Paulsen, M. A. Novak, R. Ferre, D. Mailly, R. Sessoli, A. Caneschi, K. Hasselbach, A. Benoit, and L. Thomas, J. Magn. Magn. Mater. **140–141**, 1825 (1995).

[23] C. Delfs, D. Gatteschi, L. Pardi, R. Sessoli, K. Wieghardt, and D. Hanke, Inorg. Chem. **32**, 3099 (1993).

[24 ] *For Mn12*, J.R. Friedman, M.P. Sarachik, J. Tejada, and R. Ziolo, Phys. Rev. Lett. **76**, 3830 (1996); L. Thomas, F. Lionti, R. Ballou, D. Gatteschi, R. Sessoli, and B. Barbara, Nature (London) 383, 145 (1996) ; *for Fe8*, C. Sangregorio, T. Ohm, C. Paulsen, R. Sessoli, and D. Gatteschi, Phys. Rev. Lett. **78**, 4645 (1997); W. Wernsdorfer, A. Caneschi, R. Sessoli, D. Gatteschi, A. Cornia, V. Villar, and C. Paulsen, Phys. Rev. Lett. **84**, 2965 (2000).

[25] Y. Furukawa, K. Kumagai, A. Lascialfari, S. Aldrovandi, F. Borsa, R. Sessoli and D. Gatteschi, Phys. Rev. **B 64**, 094439 (2001).

[26] A.Lascialfari, P.Carretta, D.Gatteschi, C.Sangregorio, J.S.Lord e C.A.Scott, Physica B **289-290**, 110 (2000) ; D.Gatteschi, P.Carretta e A.Lascialfari, Physica B **289-290**, 94 (2000).

[27] A.Lascialfari, Z.H.Jang, F.Borsa, P.Carretta and D.Gatteschi, Phys. Rev. Lett. **81**,3773 (1998).

[28] M.N. Leuenberger and D. Loss, Phys. Rev. **B 61**, 1286 (2000).

[29] J. Villain, F. Hartmann-Boutron, R. Sessoli, and A. Rettori, Europhys. Lett. **27**, 537 (1994); F. Hartmann-Boutron, P. Politi, and J. Villain, Int. J. Mod. Phys. **10**, 2577 (1996).

[30] T. Goto, T. Koshiba, T. Kubo, and K. Awaga, Phys. Rev. B **67**, 104408 (2003).

[31] S. Yamamoto and T. Nakanishi, Phys. Rev. Lett. **89**, 157603 (2002); H.Hori and S.Yamamoto, Phys.Rev. B 68,054409 (2003).

[32] S. Maegawa and M.Ueda, Physica B 329, 1144( 2003); T.Yamasaki, M.Ueda and S. Maegawa, Physica B 329, 1187 (2003).

[33] B.J.Suh, D.Procissi, P.Kogerler, E.Micotti, A.Lascialfari and F.Borsa, J.Mag. Mag. Mater., in press (ICM 2003).
[34] A.Lascialfari, Z.H.Jang, F.Borsa, A.Cornia, D.Rovai, A.Caneschi and P.Carretta, Phys.Rev.**B 61**, 6839, (2000).
[35] A. Cornia, et al. Phys. Rev. **B 60**, 12177 (1999).
[36] F. Meier and D. Loss, Phys. Rev. Lett. 23 5373 (1999); O. Waldmann, Europhys. Lett. **60**, 302 (2002).; A. Chiolero and D. Loss, Phys. Rev. Lett. **80**, 169 (1998).
[37] For NMR studies, see M. Chiba et al., J. Phys. Soc. Jpn. **57**, 3178 (1988); G. Chaboussant et al., Phys. Rev. Lett. **80**, 2713 (1998); M. Chiba et al., Physica (Amsterdam): **246B & 247B**, 576 (1998).
[38] A. Abragam and B. Bleaney, "Electron Paramagnetic Resonance of Transition Ions", Clarendon Press (Oxford, 1970).
[39] M.H. Julien, Z.H. Jang, A. Lascialfari, F. Borsa, H. Horvatic, A. Caneschi and D. Gatteschi, Phys.Rev.Lett. **83**, 227 (1999).
[40] M. Affronte, A. Cornia, A. Lascialfari, F. Borsa, D. Gatteschi, J. Hinderer, M. Horvatic, A.G.M. Jansen, and M.-H. Julien, Phys. Rev. Lett. **88**,167201 (2002).
[41] A. Lascialfari, F. Borsa, M-H. Julien, E. Micotti, Y. Furukawa, Z.H. Jang, A. Cornia, D. Gatteschi, M. Horvatic and J.Van Slageren, J.Magn.Magn. Mater., in press (ICM2003, invited paper).
[42] A. Cornia ,A.Fort, M.G.Pini and A.Rettori, Europhys. Lett. **50**, 88 (2000).
[43] H. Nakano and S. Miyashita, J. Phys. Soc. Jpn., **71**, 2580 (2002).
[44] F. Cinti, M. Affronte and A.G.M. Jansen, Eur. Phys. J. **B 30**, 461 (2002).
[45] M.Affronte et al. Phys. Rev. B68,104403 ( 2003).
[46] T. Goto, T. Kubo, T. Koshiba, Y. Fuji, A. Oyamada, J. Arai, T. Takeda and K. Awaga, Physica B **284-288**, 1277 (2000).
[47] Y. Furukawa, K. Watanabe, K. Kumagai, F. Borsa and D. Gatteschi, Phys. Rev. **B 64**, 104401 (2001).
[48] T. Kubo, T. Goto, T. Koshiba, K. Takeda, K. Agawa, Phys. Rev. B 65, 224425 (2002) ; T.Koshiba, T. Goto , T.Kubo and K.Awaga, Progr. Theor. Phys. ( Suppl.) **145**, 394 (2002)
[49] S.H. Baek, S. Kawakami, Y. Furukawa, B.J. Suh, F. Borsa, K. Kumagai, and A. Cornia, J. Magn. Magn. Mater., in press (ICM2003).
[50] Y. Furukawa, S. Kawakami, K. Kumagai, S-H. Baek and F. Borsa, Phys. Rev. **B68**, 180405(R) (2003).
[51] Y. Furukawa, K. Aizawa, K. Kumagai, A. Lascialfari, S. Aldrovandi, F. Borsa, R. Sessoli, D. Gatteschi, Mol. Cryst. Liq. Cryst. , **379**, 191 ( 2002) ( Conference Proceedings).
[52] R.A. Robinson, P.J. Brown, D.N. Argyriou, D.N. Hendrickson, S.M.J. Aubin, J.Phys.:Cond.Matter **12**, 2805 (2000) ; M. Hennion , L. Pardi , I. Mirebeau , E. Suard , R. Sessoli , A. Caneschi, Phys. Rev. **B56**, 8819 (1997) ; I. Mirebeau , M. Hennion , H. Casalta , H. Andres , H.U. Gudel , A.V. Irodova, A. Caneschi, Phys.Rev.Lett. **83**, 628 (1999)  ; A. Cornia , R. Sessoli , L. Sorace, D. Gatteschi , A.L. Barra , C. Daiguebonne, Phys. Rev. Lett. **89**, 257201 (2002).
[53] Y. Furukawa, K. Watanabe, K. Kumagai, F. Borsa, T. Sasaki, N. Kobayashi and D. Gatteschi, Phys. Rev. **B 67**, 064426 (2003); K. Watanabe, Y. Furukawa, K. Kumagai, F. Borsa, D. Gatteschi, Mol. Cryst. Liq. Cryst. **379**, 185 (2002).

[54] Y. Pontillon , A. Caneschi , D. Gatteschi , R. Sessoli , E. Ressouche , J. Schweizer , E. Lelievre-Berna, J.Am.Chem. Soc. **121**, 5342 (1999) ; R. Caciuffo , G. Amoretti , A. Murani , R. Sessoli , A. Caneschi , D. Gatteschi, Phys.Rev.Lett. **81**, 4744 (1998).

[55] For example, D. Gatteschi and R. Sessoli, "Magnetism: Molecules to Materials III" edited by J. S. Miller and M. Drillon (Wiley-VHC, Verlag GmbH, Weinheim, 2002).

[56] I.Tupitsyn and B.Barbara, "Magnetism: Molecules to Materials III" edited by J. S. Miller and M. Drillon (Wiley-VHC, Verlag GmbH, Weinheim, 2002).

[57] Z.H. Jang, A. Lascialfari, F. Borsa and D. Gatteschi, Phys. Rev. Lett. **84**, 2977, (2000).

[58] Y. Furukawa, W. Watanabe, K. Kumagai, Z.H. Jang, A. Lascialfari,F. Borsa and D.Gatteschi, Phys. Rev. **B62**, 14246 (2000).

[59] Y. Furukawa, K. Watanabe, K. Kumagai, F. Borsa and D. Gatteschi, Physica B **329-333**, 1146 (2003).

[60] T.Kubo, H.Doi, B.Imanari, T.Goto, K.Takeda and K.Awaga, Physica B **329-333**, 1172 (2003).

[61] T. Goto, T. Koshiba, A. Oyamada, T. Kubo, Y. Suzuki, K. Awaga, B. Barbara and J.P. Boucher, Physica B **329-333**, 1185 (2003).

[62] for Fe8 see : M. Ueda, S. Maegawa and S. Kitagawa, Phys. Rev. **B 66**, 073309 (2002) ; for Mn12 see : A.Morello, O.N.Bakharev,H.B.Brom and L.J.de Jongh, Polyhedron **22**, 1745 ( 2003).

[63] For example, M. Luis, F.L. Mettes and L. Jos de Jongh, "Magnetism: Molecules to Materials III" edited by J. S. Miller and M. Drillon (Wiley-VHC, Verlag GmbH, Weinheim, 2002).

[64] Y. Furukawa, K. Aizawa, K. Kumagai, A. Lascialfari and F. Borsa, J.Magn.Magn.Mater., in press (ICM2003); Y. Furukawa, K. Aizawa, K. Kumagai, R. Ullu, A. Lascialfari and F. Borsa, Phys. Rev. B, in press ; for proton NMR study on oriented powders see also Y.Furukawa, K. Aizawa, K.Kumagai, R.Ullu, A.Lascialfari and F.Borsa, J. Appl. Phys. **93**,7813 (2003).

[65] A. Lascialfari, F. Tabak, G.L Abbati, F. Borsa, M. Corti and D. Gatteschi, J.Appl. Phys. **85**, 4539 (1999).

[66] A.L. Barra , A. Caneschi , A. Cornia , F.F. de Biani , D. Gatteschi , C. Sangregorio, R. Sessoli , L. Sorace, J.Am.Chem.Soc. **121**, 5302 (1999) ; G. Amoretti , S. Carretta , R. Caciuffo , H. Casalta , A. Cornia , M. Affronte , D. Gatteschi , Phys.Rev. **B6410**, 104403 (2001).

[67] D. Procissi, B.J. Suh , A. Lascialfari, F. Borsa, A. Caneschi and A. Cornia, J. Appl. Phys. **91**,7173 (2002).

[68] A. Lascialfari, D. Gatteschi, F. Borsa, A. Shastri, Z.H. Jang, and P. Carretta, Phys. Rev. **B 57**, 514 (1998).

[69] R.M.Achey, P.L.Kuhns , A.P. Reyes, W.G. Moulton, N.S. Dalal, Solid State Comm. **121**, 107 ( 2002) ; Polyhedron **20**, 11, (2001) ; Phys. Rev. **B 64**, 064420 ( 2001).

[70] D.Arcon, J.Dolinsek, T.Apih, R.Blinc, N.S.Dalal, R.M.Achey, Phys. Rev. **B 58**, R2941 (1998).

[71] R.Blinc, B. Zalar, A. Gregorovic, D. Arcon, Z. Kutnjak, C. Filipic, A. Levstik, R.M. Achey, N.S. Dalal, Phys. Rev. **B 67**, 094401 (2003).

[72] M. Axenovich and M. Luban, Phys. Rev. **B 63**, 100407 (2001).

[73] J.K. Jung, D. Procissi, R. Vincent, B.J. Suh , F. Borsa, C. Schroder, M. Luban, P. Kogerelr and A. Muller, J.Appl. Phys. **91**,7388 (2002).

[74] E.Micotti, D.Procissi, A.Lascialfari,P.Carretta, P.Kogerler, F.Borsa, M.Luban, C.Baines, J.Magn. Magn. Mater., in press (ICM2003).

[75] "Biomineralization : chemical and biochemical perspectives" edited by Stephen Mann, John Webb, Robert J.P.Williams, VCH, (1989).

[76] M-E.Y. Mohie-Eldin, R.B. Frankel and L. Gunter, J.Magn. Magn. Mater. **135**, 65 (1993).

[77] J. Tejada, X.X. Zhang, E. del Barco and J.M. Hernández, Phys. Rev. Lett. **79**, 1754 (1997).

[78] J.R. Friedman, U. Voskoboynik, and M. P. Sarachik, Phys. Rev. **B56**, 10793 (1997).

[79] D. D. Awschalom, J. F. Smyth, G. Grinstein, D. P. DiVincenzo, and D. Loss, Phys. Rev. Lett. **68**, 3092 (1992).

[80] Z.H. Jang, B.J. Suh, A. Lascialfari, R. Sessoli, and F. Borsa, J. Appl. Magn. Res. **19**, 557 (2000).

[81] E.Rentschler, D.Gatteschi, A.Cornia, A.C.Fabretti, A-L.Barra, O.I.Shchegolikhina and A.A.Zhdanov, Inorg.Chem.**35**, 4427 (1996).

[82] Y. Furukawa, A. Lascialfari, Z.H. Jang , and F. Borsa, J. Appl. Phys. **87**, 6265 (2000).

[83] Y. Furukawa, M. Luban, F. Borsa, D.C. Johnston, A.V. Mahajan, L.L. Miller, D. Mentrup, J. Schnack and A. Bino, Phys. Rev. **B6**1, 8635 (2000).

[84] M. Luban, F, Borsa, S. Budko, P. Canfield, S. Jun, J.K. Jung, P. Kogerler, D. Mentrup, A. Muller, R. Modler, D. Procissi, B.J. Suh, M. Torikachvilly, Phys. Rev. **B 66**,054407 (2002).

[85] D. Procissi, B.J. Suh, J.K. Jung, P. Kogerler, R. Vincent and F. Borsa, J. Appl. Phys. **93**, 7810 (2003).

[86] J.K. Jung, D. Procissi, Z.H. Jang B.J. Suh , F. Borsa, M. Luban, P. Kogerler and A. Muller, J. Appl. Phys. **91**, 7391 (2002).

[87] B.J. Suh, D. Procissi, K.J. Jung, S. Budko, W.S. Jeon, Y.J. Kim and D.Y. Jung, J. Appl. Phys. **93**, 7098 (2003).

[88] M.Fardis, G. Diamantopoulos, M. Karayianni, G. Papavassiliou, V. Tangoulis, A. Konsta, Phys. Rev. **B 65**, 014412 (2001) .